\documentclass[twocolumn]{aa}

\usepackage{graphicx}
\usepackage{natbib}
\usepackage{times}
\usepackage{amsmath}

\begin{document}
\title{Estimation of the detectability of optical orphan afterglows}
\author{Y. C. Zou \inst{1}, X. F. Wu \inst{2}, \and Z. G. Dai \inst{1}}

\institute{Department of Astronomy, Nanjing University, Nanjing 210093, China \\
\and  Purple Mountain Observatory, Chinese Academy of Sciences, Nanjing 210008, China \\
Joint Center for Particle, Nuclear Physics and Cosmology (J-CPNPC) between
Nanjing University and Purple Mountain Observatory, Nanjing 210008, China \\
\email{ zouyc@hust.edu.cn (YCZ); xfwu@pmo.ac.cn (XFW); dzg@nju.edu.cn (ZGD)} }

\abstract
{}%context
{By neglecting sideways expansion of gamma-ray burst (GRB) jets and assuming their
half-opening angle distribution, we estimate the detectability
of orphan optical afterglows.}%aim
{This estimation is carried out by calculating the durations of off-axis optical
afterglows whose flux density exceeds a certain observational limit.}%method
{We show that the former assumption leads to more detectable orphans,
while the latter suppresses the detectability strongly compared with
the model with half-opening angle $\theta_j=0.1$. We also considered
the effects of other parameters, and find that the effects of the
ejecta energy $E_j$ and post-jet-break temporal index $-\alpha_2$
are important but that the effects of the electron-energy distribution
index $p$, electron energy equipartition factor $\epsilon_e$, and
environment density $n$ are insignificant. If $E_j$ and
$\alpha_2$ are determined by other methods, one can constrain the
half-opening angle distribution of jets by observing orphan
afterglows. Adopting a set of ``standard" parameters, the 
detectable rate of orphan afterglows is about $1.3\times 10^{-2}
{\rm deg}^{-2} {\rm yr}^{-1}$, if the observed limiting magnitude is
20 in R-band.
}%results
{}%conclusion

\authorrunning{Y. C. Zou, X. F. Wu \& Z. G. Dai}
\titlerunning{detectability of orphan afterglows}

\maketitle

\section{Introduction}\label{sec_intro}
Orphan afterglows are defined as afterglows whose gamma-ray bursts
(GRBs) are not detected, possibly because of the Doppler effect for
an off-axis observer. {If the GRB afterglows are modelled perfectly,
} the observed rate of orphans and GRBs can be used to constrain the
beaming factor of GRBs, as first proposed by \citet{r97}. However,
as many parameters are not determined well for different
afterglows, it will be difficult to constrain the beaming factor
tightly from optical orphans. Because late radio afterglows behave
 isotropic emission, the survey of radio afterglows may be helpful
for estimating the beaming factor \citep{lo02, go05}, although one
should be careful to rule out radio transients from other sources.

\begin{figure}
  \includegraphics[angle=270,width=0.5\textwidth]{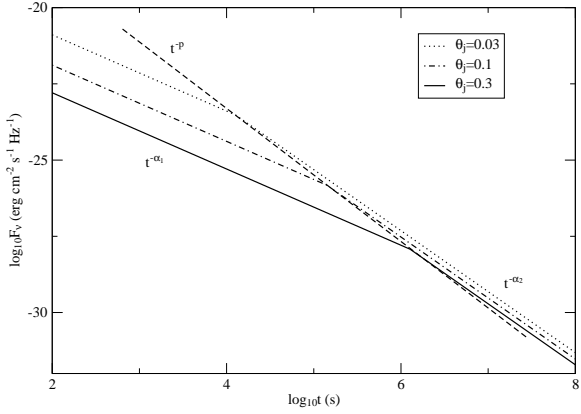}
  \caption{Sketch of afterglow light curves { from jets} without sideways expansion
  for  an on-axis observer. Dotted, { dot-dashed} and solid lines correspond
  to three jet opening angles 0.03, 0.1 and 0.3 respectively, with the same total
  kinetic energy $E_j = 1\times 10^{51}$ ergs. The dashed line is the connection
  of jet breaks.}
  \label{f_nu_t_theta}
\end{figure}

Other than constraining the beaming factor \citep{r97, dgp02, tp02},
some authors have focused on investigations of the detectability of
orphan optical afterglows both theoretically \citep{npg02, tp02} and
experimentally \citep{h04, bw04, ra05, ma06, rgs06}. \citet{bw04}
give the results of a 5-year (1999-2004) survey of optical
transients, but none was identified as an orphan. \citet{ra05}
performed a 1.5-year survey (2003 September to 2005 March) of
untriggered GRB afterglows. Although no orphan afterglow has 
been observed yet, they give the upper limit of the observed rate for a
certain limiting magnitude. The surveys are still
going on\citep{ma06,mab06}. On the theoretical side, \citet{npg02}
considered the following afterglow model: all jets propagating in a
uniform medium (ISM) have a constant initial half-opening angle
$\theta_j$ and a constant jet energy $E_j$. After a jet break takes
place because the hydrodynamics of a sideways-expansion jet
enters an exponential regime \citep{r99,sp99}, the temporal index
of light curves becomes $-p$ (where $p$ is the power-law index of
shock-accelerated electrons). This decline is too steep for  most
of the observed late afterglows\citep{lz05}. On the other hand, many
works \citep{msb00, hgd00, wl00, s03, kg03, gk03, cgv04} show
that the sideways expansion of jets is insignificant at the
relativistic stage. Thus, we consider relativistic jets without
sideways expansion, and their afterglow light curves for an on-axis
observer are shown in Fig. \ref{f_nu_t_theta}. The light curves
are shown in the spherical case and the flux density $f_\nu \propto
t^{-\alpha_1}$ {when the Lorentz factor $\gamma > 1/\theta_j$}.
After the jet break time, the light curves steepen { as $f_\nu
\propto t^{-\alpha_2}\,(\alpha_2>\alpha_1)$} because of the edge
effect\citep{mr99}, which is flatter than the sideways-expansion
case. This will lead to more detectable orphan afterglows. A
relationship between the jet break time and flux density
($f_{\nu,j}\propto t_j^{-p}$) was found by \citet{wdl04}
analytically and statistically.

Fig. \ref{f_nu_t_theta} is plotted based on the fact that the
kinetic energies of all GRB jets have a similar value
\citep{fks01,bf03}. The statistical standard energy of jets has been
discussed by several authors, e.g., \citet{lpp01}, and \citet{pk02},
who conclude that the collimation-corrected gamma-ray energy has a
relatively narrow distribution, around $5\times 10^{50}$ ergs.
\citet{bkf03} also obtained a standard kinetic energy reservoir of
afterglows from the statistics on X-ray luminosity.

Recently, the distribution of the half-opening angle or viewing
angle was investigated based on several structured-jet models
\citep{psf03, lwd04, ngg04, gpw05}. Considering a uniform,
sharp-edge jet \citep[favored by][]{ldg05}, whose light curves are
similar to those in  the universal structured jet models
\citep{rlr02}; and using the observed distribution of half-opening
angles of the jets given by \citet{ldg05}, one can derive the
intrinsic distribution of $\theta_j$, i.e., $P(\theta_j) \propto
\theta_j^{-1}$ \citep[see also ][]{xwd05}. We use this distribution
as a weight for different opening angles as suggested by
\citet{gpw05}.

By considering both the effects of the constant half-opening angle {
during} jet propagation and the distribution of {initial jet
half-opening angles}, we here estimate the detectability of orphan
afterglows and find that our results are different from the ones in
earlier works \citep[e.g.][]{npg02, tp02}. We { present} the
theoretical model in \S \ref{sec:analysis} and { give the results
of} the detectability in \S \ref{sec:results}. We summarize our
findings and present a brief discussion in \S \ref{sec:discuss}.

\section{Theoretical analysis}\label{sec:analysis}
We consider an adiabatic jet with a total { kinetic} energy $E_j$
and a half-opening angle $\theta_j$ and neglect sideways expansion.
The { hydrodynamics} of the jet behaves as a spherical case
\citep{spn98}. The Lorentz factor of the jet is given by
\begin{equation}
  \gamma(t_\oplus) = 8.9 (1+z)^{3/8} E_{j,51}^{1/8} n_0^{-1/8} \theta_{j,-1}^{-1/4}
  t_{\oplus,d}^{-3/8},
  \label{eq:gamma_t}
\end{equation}
where $z$ is the redshift of the GRB, $n$ the number density of  the
interstellar medium (ISM), and $t_{\oplus,d}$ the observed time in units of
days. We adopt the conventional notation $Q=Q_k\times10^{k}$ in this paper
except for special explanations.

For an on-axis observer, there is a break in the light curve because of the
edge effect \citep{mr99} when the bulk Lorentz factor $\gamma$  equals
$\theta_j^{-1}$. The jet-break time is given by
\begin{equation}
  t_{j}=0.82(1+z)E_{j,51}^{1/3}n_0^{-1/3}\theta_{j,-1}^2\,\,{\rm{days.}}
  \label{eq:t_j}
\end{equation}
At the jet-break time, the flux density in the slow-cooling case
($\nu_m<\nu<\nu_c$) is \citep{wdl04}
\begin{eqnarray}
  F_{\nu,j}&=& 515 t_{j,\rm{day}}^{-p}\times50.2^{2.2-p}\kappa_{f}
  \kappa_{m}^{(p-1)/2}\epsilon_{e,-1}^{p-1} \nonumber \\
& & \times \epsilon_{B,-3}^{(p+1)/4}\zeta_{1/6}^{p-1}n_0^{(3-p)/12}
E_{j,51}^{(p+3)/3}D_{L,28}^{-2} \nonumber \\
& & \times (1+z)^{(p+3)/2}(\frac{\nu}{\nu_R})^{-(p-1)/2}\,\,{\rm{\mu Jy}},
\label{eq:f_nu_j_slow}
\end{eqnarray}
where $\kappa_{m}=0.73(p-0.67)$, $\kappa_{f}=0.09(p+0.14)$, and
$\kappa_{c}=(p-0.46)\exp{(3.16-1.16p)}$ are the correction factors
\citep{gs02}; $\epsilon_e$ and $\epsilon_B$ are the energy
equipartition factors of the electrons and magnetic field, respectively;
$\zeta_{1/6}=6(p-2)/(p-1)$; $D_{L}$ is the luminosity distance; and
$\nu_{R}=4.55\times10^{14}$ Hz is the $R$-band frequency taken as
the observed frequency. On the other hand, in the fast cooling case
($\nu_c<\nu$), the flux density is \citep{wdl04}
\begin{eqnarray}
 F_{\nu,j}&=& 3508 t_{j,\rm{day}}^{-p}\times50.2^{2.2-p}\kappa_{f}
 \kappa_{m}^{(p-1)/2}\kappa_{c}^{1/2}D_{L,28}^{-2} \nonumber \\
& & \times \epsilon_{e,-1}^{p-1}\epsilon_{B,-3}^{(p-2)/4}\zeta_{1/6}^{p-1}
E_{j,51}^{(p+2)/3}n_0^{-(p+2)/12} \nonumber \\
& & \times (1+z)^{(p+2)/2}(1+Y_{j})^{-1}(\frac{\nu}{\nu_R})^{-p/2}\,\,{\rm{\mu Jy}},
\label{eq:f_nu_j_quick}
\end{eqnarray}
where $Y_j=Y(t_j)=(-1+\sqrt{1+4\eta\epsilon_e/\epsilon_B})/2$ is the Compton
parameter \citep{pk00, se01}, and $\eta$ is the radiation efficiency of
electrons. Even though most GRB afterglows match the slow cooling case in the
statistics { by} \citet{wdl04}, we here consider both fast and slow cooling
cases, which are different from \citet{npg02}, who only considered the case
$\nu>\nu_c>\nu_m$ for simplicity.

{ For an on-axis observer,} the Lorentz factor of the jet
$\gamma(t_\oplus)$ at earlier times ($t_{\oplus}<t_j$) is greater
than $\theta_j^{-1}$, so the emission properties are the same as
those from an isotropic fireball. The temporal decay index of the
flux density $F_{\nu,0}(t_\oplus)$ is $(2-3p)/4$ in the fast cooling
case and $3(1-p)/4$ in the slow cooling one \citep{spn98}. As 
index $p$ is { mainly} in the range of $2.0\sim 2.4$, we find that {
the range of the temporal index is about $-0.7$ to $-1.3$ for both
cases}, which is set to be a parameter $-\alpha_1$. When
$t_{\oplus}>t_j$ { and if $\theta_j \ll 1$ and $\gamma \gg 1$,} the
on-axis observer can only detect a fraction $\theta_j^2\gamma^2$ of
{ the} flux density in the isotropic fireball case. As
$\gamma(t_\oplus) \propto t_\oplus^{-3/8}$ { [see Eq.
(\ref{eq:gamma_t})]}, the { late} decay index of the flux density
$\alpha_2 = \alpha_1+3/4$.

%\iftrue
\iffalse Taking into account the broken power-laws of the spectrum {
for an off-axis observer with observing angle $\theta_{obs}$ in the
point source approximation that is good enough when
$\theta_{obs}>\theta_j$ \citep{gp02}, we have the flux density
\citep{Mirabel99}
\begin{equation}
  F_{\nu,\theta_{obs}} = F_{\nu,0} \left\{
  \begin{array}{ll}
  \left( \frac{\mathcal{D}_{\theta}}{\mathcal{D}_0} \right)^{3-\alpha+(p-1)/2},&
  \nu<\frac{\mathcal{D}_\theta}{\mathcal{D}_{0}}\nu_c\\
  \left(  \frac{\mathcal{D}_{\theta}}{\mathcal{D}_0} \right)^{3-\alpha+p/2}
  \left( \frac{\nu}{\nu_c}\right)^{-1/2},& \frac{\mathcal{D}_\theta}
  {\mathcal{D}_{0}}\nu_c < \nu < \nu_c \\
  \left(  \frac{\mathcal{D}_{\theta}}{\mathcal{D}_0} \right)^{3-\alpha+p/2}, & \nu>\nu_c
  \end{array}
  \right.
  \label{eq:doppler_p}
\end{equation}
where $\alpha$ is the temporal index of the on-axis light curve,
$\mathcal{D}_{\theta}=1/[\gamma(1-\beta\cos\theta_{\rm obs})]$ and
$\mathcal{D}_{0}=1/[\gamma(1-\beta)]$ are the Doppler factors
corresponding to the observing angle $\theta_{\rm obs}$ and $0$
respectively, and $\beta = \sqrt{1-1/\gamma^2}$ is the velocity in
units of $c$. Note that $\nu_c$ is the value in the on-axis
observer's frame. }\fi

For an off-axis observer with observing angle $\theta_{obs}$, {the
time and frequency from on-axis ($t_0, \nu_0$) and off-axis ($t,
\nu$) jets satisfy $t_0/t \simeq \nu/\nu_0 =
(1-\beta)/(1-\beta\cos{\theta_{\rm obs}}) \equiv a$,
where %$\theta=\theta_{\rm obs}$,
 $\beta = \sqrt{1-1/\gamma^2}$ is the velocity in units of $c$;
thus the flux density is
\begin{equation}
  F_{\nu}(\theta_{obs},t) = a^3 F_{\nu/a}(0,at),
  \label{eq:doppler}
\end{equation}
in the point source approximation, which is
good enough when $\theta_{obs}>\theta_j$ \citep{gp02}. }

\begin{figure}
  \includegraphics[angle=270,width=0.5\textwidth]{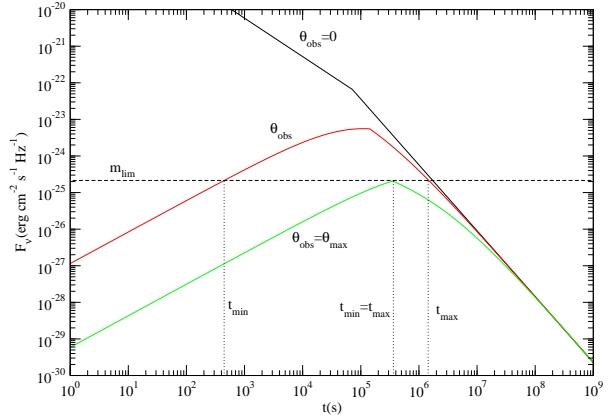}
  \caption{Sketch for observations of orphan R-band afterglows. The three solid
  lines indicate the light curves with different observing angles. { The}
  one with $\theta_{obs}=0$ is not an orphan afterglow, which is plotted
  here as a reference. The { horizontal} dashed line represents { a given}
  limiting magnitude. The earliest and latest times ($t_{\min}$ and $t_{\max}$)
  at which
  an orphan afterglow are observed are { represented by vertical} dotted lines.}
  \label{fig:f_nu_t_theta_obs}
\end{figure}

Given a limiting flux density $f_{\nu,\lim}$ (corresponding to a {
limiting} magnitude $m_{\lim}$) { for an instrument with a fixed
exposure time}, we can calculate the detectable duration { of an
orphan afterglow:} $t_{obs}(z, \theta_{obs}, \theta_j, m_{\lim}) =
t_{\max}-t_{\min}$, where $t_{\min}$ and $t_{\max}$ represent the {
earlier and later times} when $F_{\nu,\theta_{obs}} = F_{\nu,\lim}$.
If the maximum observed flux density $F_{\nu,\theta_{obs},\max} <
F_{\nu,\lim}$, we take $t_{obs}=0$. Fig.
\ref{fig:f_nu_t_theta_obs} shows the light curves of afterglows for
different observing angles.

Following \citet{npg02}, we assume that the GRB rate $n(z)$ is
proportional to the star formation rate (SFR), but we use { a
different SFR model as follows \citep{pm01}},
\begin{equation}
  n(z)=B \frac{e^{3.4z}}{e^{3.4z}+22} \frac{\sqrt{\Omega_m(1+z)^3+
  \Omega_{\lambda}}}{(1+z)^{1.5}}
  \label{eq:SFR}
\end{equation}
where $\Omega_m$ and $\Omega_\lambda$ are the cosmological
parameters, and $B$ the normalization factor. The parameter $B$
satisfies $R_{GRB}^{true}=\bar{f}_{b}R_{GRB}^{obs}=\int
_{0}^{10}(dV/dz)n(z)/(1+z)dz$, where {
$\bar{f}_b=\int_{\theta_{j,\min}}^{\theta_{j,\max}}(1-\cos\theta_j)^{-1}
P(\theta_j){\rm{d}}\theta_j/\int_{\theta_{j,\min}}^{\theta_{j,\max}}
P(\theta_j){\rm{d}}\theta_j$} is the mean beaming factor of GRBs and
$R_{GRB}^{obs} = 667 \mbox{yr}^{-1}$ is the observed GRB rate. Here
we assume all the GRBs can be observed if the observer is located
within the solid angles of the jets and the redshift range is
$0<z<10$.

If the exposure time is not too long (shorter than $t_{obs}$),
the number of detectable orphan afterglows in a single snapshot over
the whole sky can be expressed as
\begin{eqnarray}
  N_{orph} &=& \int_0^{10} {n(z)\over (1+z)} {{\rm{d}}V(z) \over {\rm{d}}z} {\rm{d}}z
  \int_{\theta_{j,\min}}^{\theta_{j,\max}} P(\theta_j) {\rm{d}} \theta_j \nonumber \\
 & &
  \times \int_{\theta_j}^{\theta_{\max}(z,m_{\lim})}  t_{obs}(z,\theta_{obs},
  \theta_j,m_{\lim}) \theta_{obs} {\rm{d}}\theta_{obs} \ ,
  \label{eq:N_orphan}
\end{eqnarray}
where $\theta_{\max}(z,m_{\lim})$ is the maximum observing angle,
which satisfies $t_{\max}(\theta_{\max}) = t_{\min}(\theta_{\max})$,
and $P(\theta_j)$ is the { observational} distribution function of
half-opening angles of the jets with the upper and lower limits
$\theta_{j,\max}$ and $\theta_{j,\min}$, which satisfies {
\citep{ldg05}}
\begin{equation}
       P(\theta_j)= \frac{\theta_j^{-1}}{\ln(\theta_{j,\max}/\theta_{j,\min})} .
\end{equation}

\section{Numerical results}\label{sec:results}
\begin{figure}
  \includegraphics[angle=270,width=0.5\textwidth]{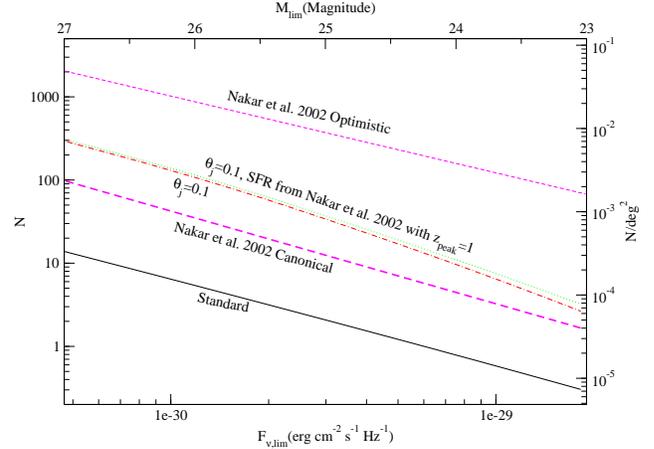}
  \caption{The estimated number of orphan afterglows in a snapshot for the whole sky,
  as a function of the limiting flux density of detectors. The solid line represents
  our standard parameterized result, with $E_j = 1\times 10^{51} {\rm erg},
  n=1 {\rm cm}^{-3}, p=2.2, \alpha_2=1.8, \epsilon_e=0.1, \epsilon_B=0.01,
  \nu = 4.55\times 10^{14}{\rm Hz}$, and a power-law distribution of half-opening
  angles of the jets with $\theta_{j,\min}=0.01$, $\theta_{j,\max}=1$.
   The thick dashed line and the thin dashed line are respectively the
  canonical line and optimistic line in \citet{npg02}, who assumed a
  laterally spreading jet. The jets have fixed initial half-opening
  angles { $\theta_j=$0.1 and 0.05}.
  The dotted line
  denotes the same parameters as the canonical line in \citet{npg02}, {but the
  assumption of no sideways expansion is used}.
  The { dot-dashed} line is the same as the dotted one except for the
  SFR model in \citet{pm01}.
%  The dashed line is the optimistic line from \citet{npg02}. The dotted line is plotted
% by considering one sole opening angle of $\theta_j=0.1$, compared with the dashed one.
  }
  \label{fig:compare}
\end{figure}

If the model parameters are given, the detectability { of orphan
afterglows} can be estimated by Eq. (\ref{eq:N_orphan}). The
main difference in detectability comes from the limiting magnitudes
of detectors. Fig. \ref{fig:compare} shows the number of orphan
afterglows that can be detected by one exposure on the whole sky.
The solid line is the standard result, with parameters $E_j =
1\times 10^{51} {\rm erg}, n=1 {\rm cm}^{-3}, p=2.2, \alpha_2=1.8,
\epsilon_e=0.1, \epsilon_B=0.01, \nu = 4.55\times 10^{14}{\rm Hz}$,
$\theta_{j,\min}=0.01$, and $\theta_{j,\max}=1$, where  $E_j, n,
p, \epsilon_e, \epsilon_B, {\rm and}~ \nu$ are the same as in
\citet{npg02}. As the pre-break temporal index of the optical light
curve is about $-1$ \citep{zm04}, we choose $\alpha_2 =
\alpha_1+3/4\simeq 1.8$. The $\theta_{\min}=0.01$ is adopted from
\citet{ldg05}. We take $\theta_{\max}=1$, which does not influence {
the estimation significantly} when we consider the distribution of
half-opening angles of the jets. We also show the results of
\citet{npg02} in this figure with their canonical { (
 $\theta_j=0.1$)} and optimistic { (
$\theta_j=0.05$)} parameters. For comparison, we plot the dotted
line for a fixed half-opening angle $\theta_j=0.1$ and the SFR model
in \citet{npg02}. One difference between the dotted line and the thick
dashed line is the temporal index after the break time. We can see
that the difference in detectability is due to the sideways expansion.
Approximately, the flux density after the break time $t_j$ is
$F_\nu(t_{\oplus}) \simeq F_{\nu,j} ({{t_{\oplus}}/{t_j}})^{-p}$ in
the sideways expansion case and $F_\nu(t_\oplus) \simeq F_{\nu,j}
({{t_{\oplus}}/{t_j}})^{-\alpha_2}$ { in the non-sideways expansion
case}. Note that $F_{\nu,j}$ and $t_j$ have the same values in both
cases, since the breaks both take place when $\gamma \simeq
1/\theta_j$, and before the jet break time, both cases show
isotropic evolutional behavior \citep{r99, mr99}. Neglecting
$t_{\min}$ and $\theta_j$ in Eq. (\ref{eq:N_orphan}) and the {
potential} influence of { different spectra}, we obtain the ratio of
{the detectabilities in} the two cases (i.e., no sideways expansion
vs. sideways expansion): $N_{orph, NSE}/N_{orph, SE} \simeq
(F_{\nu,j}/F_{\nu,\lim})^{11/(8\alpha_2)-3/(2p)}$. In general, if
$\alpha_2<p$, { then $11/(8\alpha_2) > 3/(2p)$ and thus $N_{orph,
NSE}>N_{orph, SE}$}. For a larger limiting magnitude (i.e. smaller
$F_{\nu,\lim}$), the ratio becomes higher. This is why in Fig.
\ref{fig:compare} the dotted line is higher than the { thick} dashed
line for greater $m_{\lim}$. To show the effect of different SFR
models,  the dot-dashed line uses the SFR model in \citet{npg02}
\footnote{  The form of the SFR is
\begin{equation}
  n(z) = B \left\{
  \begin{array}{ll}
   10^{0.75z} & z \leq z_{\rm peak}, \\
   10^{0.75z_{\rm peak}} & z>z_{\rm peak},
  \end{array}
  \right.
  \nonumber
\end{equation}
where $B$ is the normalization factor.}
 with $z_{\rm peak}=1$, and the dotted line considers the
SFR model in Eq. ( \ref{eq:SFR}). These two lines are close to each
other, which shows that the effect of the SFR { models} is
insignificant. We note that the distribution of half-opening angles
of the jets leads to further suppression of the detectability, which
results in the difference between the { dot-dashed} and the
solid lines. Combining the effects of the sideways expansion and
distribution of the jet's half-opening angles, we obtain standard
results (solid line in Fig. \ref{fig:compare}).

\begin{table}
 \caption{The ratio of the number of orphan afterglows to the total number
 of afterglows for different limiting magnitudes of detectors. The values { without
 and with} brackets correspond to models A and B
 in \citet{npg02} respectively to compare with their results.}
 \label{tab:ratio}
 \begin{tabular}{cc|ccc}
   \hline $z_{\rm peak}$ & $\theta_j$ & $m_{\rm lim}$=23 & $m_{\rm lim}$=25 & $m_{\rm lim}$=27 \\
   \hline 1            & 0.05       & 0.56(0.68)           & 0.73(0.78)           & 0.84(0.86) \\
          1            & 0.10       & 0.33(0.39)           & 0.40(0.58)           & 0.63(0.71) \\
          1            & 0.15       & 0.26(0.29)           & 0.34(0.40)           & 0.41(0.59) \\
          2            & 0.10       & 0.22(0.36)           & 0.34(0.55)           & 0.60(0.70) \\
%           &   SFR in the paper     & 0.30           & 0.38           & 0.62 \\
%           &   standard             & 0.54           & 0.60           & 0.63 \\
   \hline
 \end{tabular}
\end{table}

A more detailed comparison with \citet{npg02} was performed and the
results are listed in Table \ref{tab:ratio}, {which includes the
ratio of the numbers of observable orphan afterglows to total
observable afterglows}. We choose the same SFR model (see
Eq. (13) in Nakar, Piran \&
Granot 2002), and the same model A and model B
(i.e., { the angle in Lorentz transformation}
$\theta=\theta_{obs}$ for model A, and
$\theta=\max(0,\theta_{obs}-\theta_j)$ for model B),  and
neglect sideways expansion. We conclude that the ratios for model B
are all higher than those for model A. Our ratios are somewhat
lower than those in \citet{npg02}. This is because, { for a fixed
half-opening angle, the on-axis afterglow is brighter than in
the sideways expansion case} and the { flux density} of orphan
afterglows does not increase very much, being due to the Doppler effect
(see Eq. (\ref{eq:doppler})).

Recently, \citet{ra05} performed a search for orphan afterglows, but
none has been detected. They gave an upper limit for the observed
rate $\eta_{\max}<1.9\,\rm{deg}^{-2}\rm{yr}^{-1}$ by using the
method { suggested by} \citet{bw04}:
$\eta=N/(\langle\varepsilon\rangle E) \,\rm{events}\,
\rm{deg}^{-2}\rm{yr}^{-1}$, where $N$ is the number of detected
orphans, $E$ the exposure, and $\langle\varepsilon \rangle$ 
the efficiency. Assuming a 30-minute exposure time as in
\citet{ra05}, we obtain the exposure {$E\simeq 2.35\,
\rm{deg}^{2}\rm{yr}$ } for one whole sky survey. If the theoretical
efficiency $\langle \epsilon \rangle $ is assumed to be unity, the
observed rate is then
{$\eta \simeq 1.3\times 10^{-2}{\rm deg}^{-2} \rm{yr}^{-1}$} and the %optimistic
detectability $N$ is extrapolated to the 20th magnitude ($N \sim 0.03$)
in the standard model. This is well below the
upper limit $ 1.9\, \rm{deg}^{-2}\rm{yr}^{-1}$ estimated by \citet{ra05} for
the survey with { limiting} magnitude 20.

\begin{figure}
  \includegraphics[angle=270,width=0.5\textwidth]{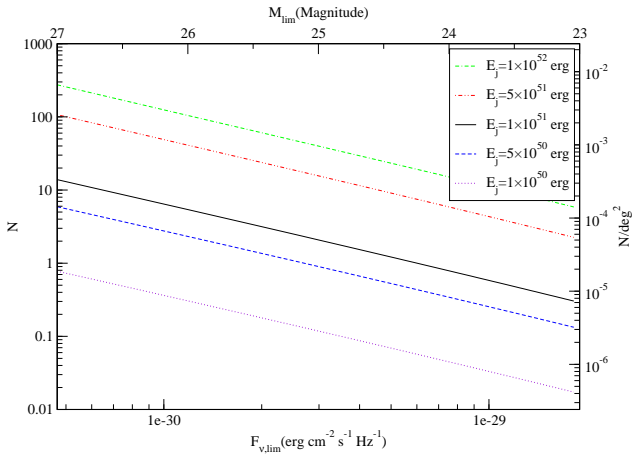}
  \caption{Same as Fig. \ref{fig:compare} with the standard
  parameters but for different total kinetic energies. From top to bottom, the $E_j$s are
  $1.0\times 10^{52}$erg, $5.0\times 10^{51}$erg, $1.0\times 10^{51}$erg,
  $5.0\times 10^{50}$erg, and $1.0\times 10^{50}$erg, while the
  solid line is the standard one.}
  \label{fig:E_j}
\end{figure}

We also calculated the detectability for different parameters to
show their effects. Fig. \ref{fig:E_j} shows the detectability
with different total kinetic energy $E_j$. The solid line is the
standard one, the same as the { solid} line in Fig. \ref{fig:compare}.
If the value of $E_j$ increases by one order of magnitude, the
detectability increases by a factor of about 12. It is reasonable
that $N$ is greater for a higher kinetic energy. Generally speaking,
the detectability is sensitive to the $E_j$.
%The survey for orphan
%afterglows may be used to constrain this parameter.

\begin{figure}
  \includegraphics[angle=270,width=0.5\textwidth]{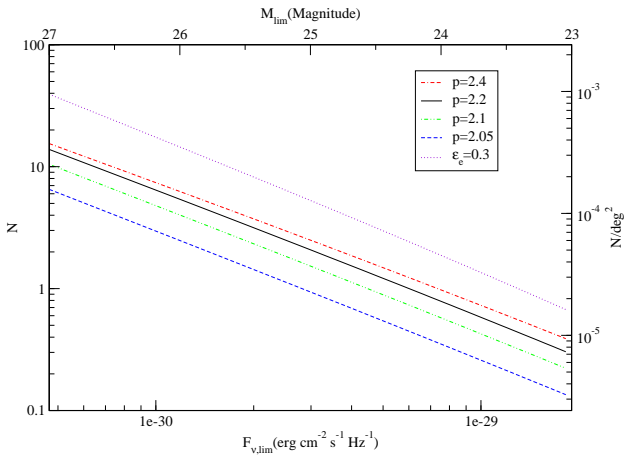}
  \caption{Same as Fig. \ref{fig:compare}. The parameters are the same
  as the standard ones (solid line) but for $p=2.4$ (the { dot-dash-dashed} line),
   $p=2.1$ (the { dot-dot-dashed} line), $p=2.05$ (the {\rm dashed} line), and
  $\epsilon_e=0.3$ (the {\rm dotted} line).}
  \label{fig:p-e}
\end{figure}

Fig. \ref{fig:p-e} shows the detectability for different $p$ and
$\epsilon_e$, with the solid line still the standard one.
Remarkably, these parameters have minor effects on the results. Note
that the $p$ and $\alpha_2$ are independent parameters here, unlike
the sideways expansion case, { where} $\alpha_2=p$ \citep{gk03}, so
the variation in $p$ does not change the temporal index $\alpha_2$.
The parameter $p$ influences the flux density at time $t_j$, which
can be seen { in} Eqs. (\ref{eq:f_nu_j_slow}) and
(\ref{eq:f_nu_j_quick}). The flux density is approximately
proportional to $\epsilon_e$, and the detectability is somewhat
sensitive to $\epsilon_e$. From those two equations, we can find
that the other parameters $n$ and $\epsilon_B$ { do not}
significantly influence the detectability of optical orphan
afterglows.

\section{Discussions}\label{sec:discuss}

We have calculated the number of orphan optical {(  R-band)}
afterglows that can be detected in an ideal survey of the whole sky
with different limiting magnitudes. We considered jets without
sideways expansion during the afterglow phase. This leads to a
flatter light curve after {the jet-break time} $t_j$, compared with
the sideways-expansion case. Thus, orphan afterglows can persist
for a longer time (above a certain flux density), and more expected
orphans can be detected.
%This enlarges about one order { of magnitude} for the
%number of detectable orphan afterglows { compared with the previous
%estimation}.
The distribution of half-opening angles of jets was also considered,
which suppresses the number of optical orphans { compared with the
case that all jets have one single half-opening} angle
$\theta_j=0.1$. When combining { these} two effects, the detectability is
less than the canonical results of \citet{npg02}, { who considered
the case in which all jets have an initial half-opening angle
$\theta_j=0.1$ with sideways expansion.}

From Figs. (\ref{fig:compare})-(\ref{fig:p-e}), we can conclude
that the main factors for the detectability are the total kinetic
energy $E_j$, half-opening angle of jet $\theta_j$ and temporal
index $\alpha_2$.  As $E_j$ is considered to be a standard energy
\citep{fks01}, our results {can be} used to constrain the value of the
$E_j$ by detecting orphans. { If} $E_j$ and $\alpha_2$ are
determined accurately by other methods, the distribution function of
the jet's { half-opening} angles { can} be determined well by
observations of orphan afterglows.

However, our estimation is simplified in several aspects. First, the
parameters are undetermined and diverse { in} different bursts. For
a variable parameter, we should know its distribution. But this is
very difficult. For example, the circum-burst environment seems to
be an ISM, a wind, or another density-profile media, but these media
cannot be determined clearly in well-observed GRBs. For simplicity,
we choose the ISM with number density $n=1 \rm{cm}^{-3}$. Second,
orphan afterglows may have properties similar to other phenomena:
e.g., failed gamma-ray burst \citep{hdl02, r03, hlc05} and
``on-axis orphan afterglow'' \citep{np03}. Third, we { use} the
model with a power-law distribution of half-opening angles of
uniform jets. However, there are some other structured jet models
that cannot be ruled out \citep{dg01,rlr02,zm02}. These models also
affect the detectability. { Fourth, we assume that the GRB rate is
proportional to the SFR, so our results are SFR-dependent.} Fifth, 
when observing, one should distinguish orphan afterglows from
other transients carefully \citep{lo02, bw04, go05}. {Finally, the
dust grains within the jet's opening solid angle may be evaporated
by the prompt UV/X-ray photons, and  the dust
is possibly opaque  in optical/UV bands outside the jet cone, so an optical orphan
afterglow may be generally suppressed.}

{\it Acknowledgements.} We would like to thank the anonymous referee 
and Steven N. Shore for valuable
suggestions, and E. Nakar for helpful discussions. This work was
supported by the National Natural Science Foundation of China
(grants 10233010, 10221001, and 10503012). XFW acknowledges the support
from the China Postdoctoral Foundation, K.C.Wong Education Foundation
(Hong Kong), and Postdoctoral Research Award of Jiangsu Province.

\end{document}